\documentclass[10pt,twoside]{myhsqcd}
\usepackage{epsf,amsmath}
\usepackage{graphicx}

\newcommand{\eq}{\begin{equation}}
\newcommand{\eqx}{\end{equation}}
\newcommand{\eqn}{\begin{eqnarray}}
\newcommand{\eqnx}{\end{eqnarray}}
\newcommand{\ad}{{a^{\dagger}}}
\newcommand{\nn}{\nonumber}
\newcommand{\ra}{\rangle}
\newcommand{\la}{\langle}
\newcommand{\Ad}{{A^{\dagger}}}
\newcommand{\Qd}{{Q^{\dagger}}}
\newcommand{\fd}{{f^{\dagger}}}
\begin{document}

\title{FROM LARGE N QUANTUM MECHANICS \\
TO PLANAR QUANTUM FIELD THEORY\thanks{TPJU - 5/2008}}

\author{\underline{J. Wosiek} \\ \\
M. Smoluchowski Institute of Physics, Jagellonian University\\
Reymonta 4, 30-059 Krakow, Poland\\
E-mail: wosiek@th.if.uj.edu.pl }

\maketitle

\begin{abstract}
\noindent We review a performance of Fock space methods in
calculating spectra of a range of supersymmetric models with gauge
symmetry. Examples include: a) SU(2) Supersymmetric
Yang Mills Quantum
Mechanics in four euclidean dimensions, b) Quantum Mechanics
of one fermion and one boson with infinite number of colours,
and c) planar 1+1 dimensional Yang Mills theories with
adjoint matter. Infrared divergencies of the latter
theories with
scalars are briefly discussed and a possible dynamical
solution of the problem is suggested.
\end{abstract}

%\markright{}

%\renewcommand{\@evenhead}

\markboth{\large \sl \underline{J. Wosiek}
\hspace*{2cm} HSQCD 2008} {\large \sl \hspace*{1cm} FROM LARGE N QUANTUM
MECHANICS TO PLANAR ...}

\section{Quantum mechanics and a space reduced field theory}
It is usually believed that simple quantum mechanics
of the finite number of degrees of freedom can hardly teach us
something about the quantum field theory with all its subtleties
like renormalization, spontaneous symmetry breaking, etc.
This is in general correct, however there exists a class
of quantum mechanics which are being intensively studied
precisely because of their connection to the field theoretical
systems \cite{Bj,CH,dWLN}. Consider the following hamiltonian
$i=1,..,D-1,\;\;\; a=1,...,N^2-1. $
\eqn
H &=&  {1\over 2} p_a^ip_a^i + {g^2\over
4}f_{abc} f_{ade}x_b^i x_c^j x_d^i x_e^j + {i g
\over 2} f_{abc}\psi_a^{\dagger}\Gamma^k\psi_b x_c^k,
\label{eq:Hamiltonian}
\eqnx
It describes a quantum mechanical system with the
finite number of degrees of freedom (e.g. 15 for D=4 and N=2)
and results from the dimensional reduction of the D
dimensional supersymmetric Yang-Mills theory to one point
in the D-1 dimensional space. In spite of the reduction,
it still
has many nontrivial properties inherited from the parent,
space extended, theory \cite{Lu,vB,BFSS} and can teach
us quite a lot about the latter. Even the SU(2) model
is not soluble for D=4. We have studied it numerically
by constructing a gauge invariant basis of bosonic and
fermionic Fock states, cutting off the total number of
bosonic quanta and subsequently increasing the cutoff
\cite{JW1}. The algorithm worked very satisfactorily and
we  were able to uncover a rich, manifestly
supersymmetric
spectrum with coexisting localized and non-localized states,
SUSY vacua and a fractional bulk value of the Witten index \cite{CW1,CW2,K}.
All these results agree with \cite{ST}, and in some cases extend \cite{dWLN},
theoretical predictions.
\section{Large N quantum mechanics}
There is much interest in the large N limit of quantum systems
and above approach turns out to be very useful
in studying such models as well.
Direct
calculations allow to obtain spectra for the
first few lowest values of N. An extrapolation
to $N=\infty$  does not seem feasible, however.
 Nevertheless, it appears that one can calculate analytically
matrix elements of typical hamiltonians directly at $N=\infty$.
\cite{VW1}. We have studied with this method  a close cousin
of the space reduced $S Y M_2$ in the planar limit. Its hamiltonian
reads
$H=\{Q,\Qd\}, Q= \sqrt{2} Tr [f \ad(1+g\ad)],
 \Qd = \sqrt{2} Tr [\fd (1+g a) a]$, or explicitly
\eqn
H&=&Tr[ \ad a + g(\ad^2 a + \ad a^2) + g^2 \ad^2 a^2]\nn\\
&+& Tr[\fd f + g ( \fd f (\ad+a) + \fd (\ad+a) f) \nonumber \\
&+& g^2 ( \fd a f \ad + \fd a \ad f + \fd f \ad a + \fd \ad f a)].\nn
\eqnx
It was found that, in spite of its simplicity,
the system exhibits many interesting phenomena: unbroken
supersymmetry, the phase transition in the 't Hooft coupling,
 at $\lambda=1$, exact
duality
between the strong and weak coupling phases, rearrangement
of supermultiplets across the transition point and emergence if the new vacua in the strong
coupling phase \cite{VW1,AdG,OVW}. All this was first discovered numerically and
subsequently confirmed by the analytic solution. Moreover,
at
strong coupling the model is exactly equivalent to the XXZ Heisenberg
spin chain and, at the same time, to the lattice gas of q-bosons.
This proves existence of a hidden supersymmetry in these
well explored statistical models \cite{VW3,JW2}.
\section{Planar field theories in 1+1 dimensions}
Extension to the field theoretical systems
is in principle straightforward. One has to diagonalize
the Hamiltonian matrix calculated in the physical basis of harmonic
oscillators. The obviously crucial difference
is that now we deal with the infinite number of (e.g. momentum)
degrees
of freedom. Nevertheless one can define
cutoff schemes which allow to extract meaningful continuum
physics. There exist two popular ways to introduce a cutoff.

{\em Light Cone Discretization (LCD)} \cite{T} replaces the
total
momentum $P$ of a proton, say, by an integer $K$. This
momentum
can then be split between various numbers of partons,
each carrying,
an integer momentum $r_i > 0$. Therefore all partitions
of $K$
into sets of integers $\{r\}$ define a finite Fock space
bound
by only one cutoff, $K$, which discretizes momenta and, at the same time,
cuts the multiplicities of partons \cite{T,Br,HP}. The Hamiltonian matrix
$\la \{r\} | H | \{s\} \ra$ is then calculated and diagonalized,
similarly as in the case of quantum mechanics. The whole art consists
of the meaningful extrapolation with $K \rightarrow \infty$.

The second approach consists of solving the
{\em Integral Equations (IE)} \cite{Kl1}
which are equivalent to the eigenequation $ H|\Phi\ra = M^2|\Phi\ra $
(in the (LC) formulation a Hamiltonian
is proportional to a $mass^2$ operator). Decomposing a bound state
into its Fock components
$ |\Phi \ra = \sum_{n=2}^{\infty} \int [dx] \delta(1-x_1-x_2-\dots x_n)
\Phi_n(x_1,x_2,\dots x_n)|x_1,\dots,x_n\ra$, turns the eigenequation
into an infinite hierarchy of integral equations
\eqn
M^2 \Phi_n(x_1\dots x_n)= A \otimes \Phi_n + B
\otimes \Phi_{n-2} + C \otimes \Phi_{n+2}, \label{ie}
\eqnx
where each term is a convolution of the wave function with the
amplitude for scattering,
emission and fusion of partons respectively. Appropriate amplitudes
can be readily read from the LC form of the hamiltonian.
In practice one has to cut the hierarchy limiting
the multiplicity of partons. At fixed maximal multiplicity
$n_{max}$ the LCD is indeed the momentum-discretized version of IE
cut to $n \le n_{max}$. In Fig.1 we compare the LCD simulations
restricted to $n_{nmax}=2$ with the IE solution at the
same multiplicity. The agreement is satisfactory, however
very fine discretization is required to see the convergence
($K\sim 2000$)\footnote{In contrast
solving integral equations with two partons required only few
basis functions}.
This is due to the singular nature of the scattering
amplitude ( $A \sim P \frac{1}{x^2}$ ).  If one allows for arbitrary number
of partons in the LCD approach the rapid growth of the number
of states with $K$ excludes $ K>25 $ which makes extrapolation
to $K=\infty$ rather delicate \cite{Br,HP,Kl1}.

\begin{figure}[!thb]
\vspace*{7.0cm}
\begin{center}
\includegraphics{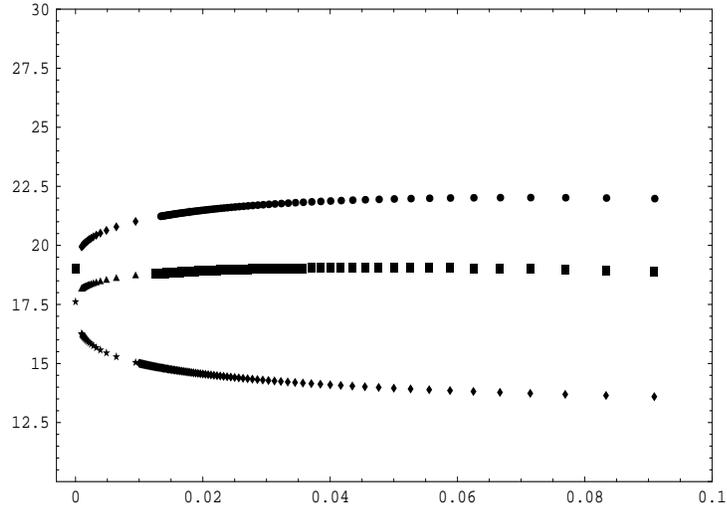}
%\centerline{\epsfxsize=2.9in\epsfbox{kim_mephi_lep.ps}}
\caption[*]{ Comparison of the LCD and IE results (symbols at $1/K=0$) }
\end{center}
\label{fig1}
\end{figure}

\section{Infrared divergencies for theories with scalars}
The previous discussion applies to a generic
LC Hamiltonians. The calculations reported
in Fig.~\ref{fig1} were done for the $YM_2$ with adjoined
fermions. There is also some interest in $YM_2$ with adjoined
scalar matter (which can be thought of as the dimensional reduction
of $YM_3$), and finally one may combine both
into
the two-dimensional supersymmetric Yang-Mills theory, $SYM_2$
\cite{MSS}. However introducing scalars brings
one more complication: the integral equations have infrared logarithmic
divergence. It appears explicitly in the mass term which is
part of the diagonal (in multiplicity) transition in
(\ref{ie}). One way to dispose of it is the
"mass renormalization" introduced in \cite{Kl1}. This might
be possible for the $YM_2$ with a scalar matter, however
for the supersymmetric model such a procedure would break
supersymmetry. The attractive possibility to maintain
SUSY would occur if the above divergence was canceled dynamically
by other contributions \cite{VW4}.

Some support for this
 idea is provided by
Fig.2 where we compare the lowest mass obtained
by the LCD simulations with (open circles) and without
(filled circles) the multiplicity cutoff. Clearly the
dependence on the cutoff is weaker (indicating possible

\begin{figure}[!thb]
\vspace*{7.0cm}
\begin{center}
\includegraphics{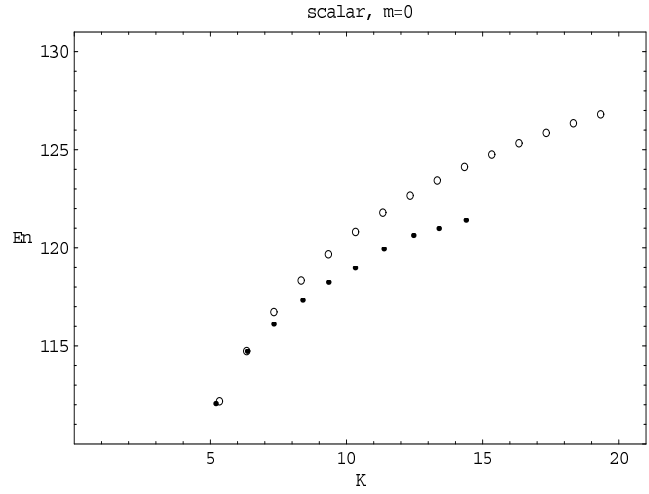}
%\centerline{\epsfxsize=2.9in\epsfbox{kim_mephi_lep.ps}}
\caption{ The LCD results without (filled)
and with the multiplicity cut (open,$n_{max}=2$).}
\end{center}
\label{fig2}
\end{figure}

\noindent convergence at larger K) when all multiplicities are allowed.
Second piece of evidence comes from careful inspection
of the integral equations (\ref{ie}) which shows that,
when the wave functions have a suitable divergence
at $x \sim 0$, additional divergences occur which in fact
cancel the original IR divergence of the mass term \cite{VW4}.
For this mechanism to work one has to include {\em all}
Fock components in eqns (\ref{ie}). This is consistent
with the message learnt from Fig.2.
In fact the whole mechanism is analogous to the classical
Bloch-Nordsieck treatment of the IR singularities in QED
as discussed in the following Section.
\section{The Bloch-Nordsieck inspired toy model}
A good insight into the mechanism of the above
dynamical cancelations is provided by the following Light Cone
model hamiltonian
\eqn
H=\int_0^P dk \left[ \ad_k a_k - j_k (\ad_k+a_k) +
 j_k^2\right] = \sum_k H_k,\;\;\;j_k=\frac{g}{\sqrt{k}}\label{hbn}\\
H_k=A_k^{\dagger} A_k,\;\;\;A_k = e^{-i P_k j_k} a_k e^{ i P_k j_k},\;\;\;
P_k=\frac{1}{i\sqrt{2}} (a_k-\ad_k).
\eqnx
The model is of course soluble, the eigenstates are the BN coherent states
\eqn
|n_k\ra_{new} = \frac{\Ad_k^n}{\sqrt{n!}} |0\ra_{new}=  e^{-i P_k j_k} |n_k\ra_{old}
\eqnx
and the eigenvalues are integer
$M^2=\sum_k  n_k$.

The integral equations (\ref{ie}) can be easily derived
\eqn
M^2 f_n(x_1,\dots,x_n) = \left( n + \int_0^1 j(x)^2 dx \right)
f_n(x_1,\dots,x_n) \nn\\
- \sum_{i=1,n} j(x_i) f_{n-1}(x_1,..,x_{j-1},x_{j+1},..,x_n)
- \int_0^1 j(x) f_{n+1}(x_1,...,x_n,x) dx \label{teq}
\eqnx
and are divergent for the sources (\ref{hbn}),
but in fact they must give (and they do!) the above
finite spectrum.

We expect that the similar cancelations occur in the case
of $YM_2$ with adjoint scalars and also in $SYM_2$.
Figure 3 lends some support for this analogy.
On the left hand we show numerical solutions of (\ref{teq})
for the first three eigenvalues as the function of the
IR cutoff $\epsilon$. Three curves for each eigenvalue
correspond to increasing multiplicities (top to bottom: $n_{max}=2,3,4$).
At finite $n_{max}$ eigenvalues are divergent at small $\epsilon$
as expected. However for increasing $n_{max}$ the divergence is
shifted towards smaller and smaller $\epsilon$ and the integer
values of $M^2_n$ are better and better approximated.
On the right hand we compare in the same way
the masses obtained from the LCD and IE for $YM_2$ with adjoint
 scalars \cite{VW4}. Both calculations were done only in
 the two-parton sector. Results are not far from each other
 showing the onset of the divergence at low $\epsilon$.
It is expected that as we include
higher Fock sectors the two curves would converge and the IR
divergence will shift to yet smaller $\epsilon$, similarly to
the toy model example.

\begin{figure}[!thb]
\vspace*{7.0cm}
\begin{center}
\includegraphics{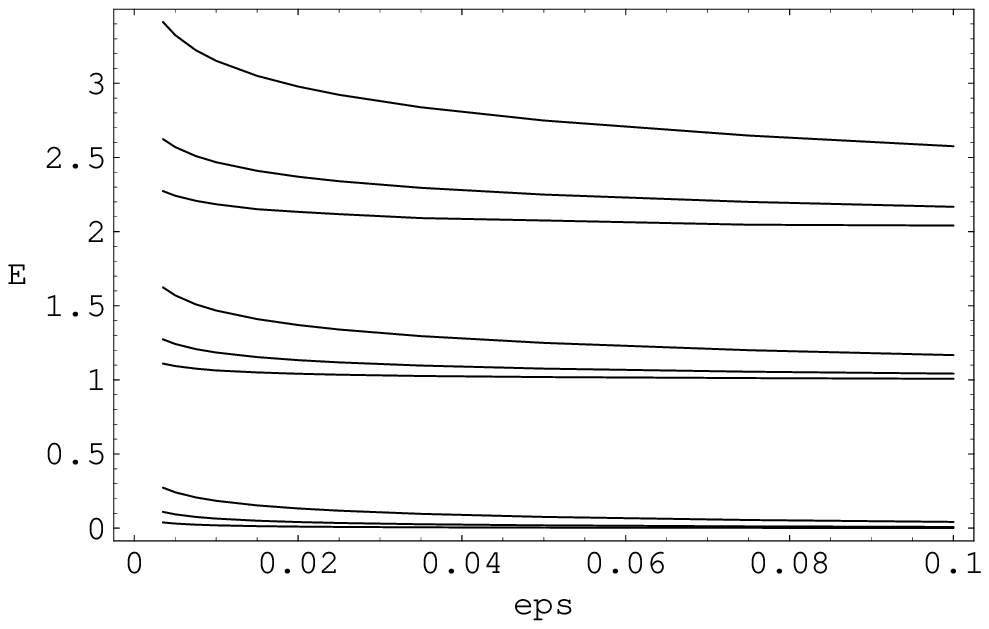}
\includegraphics{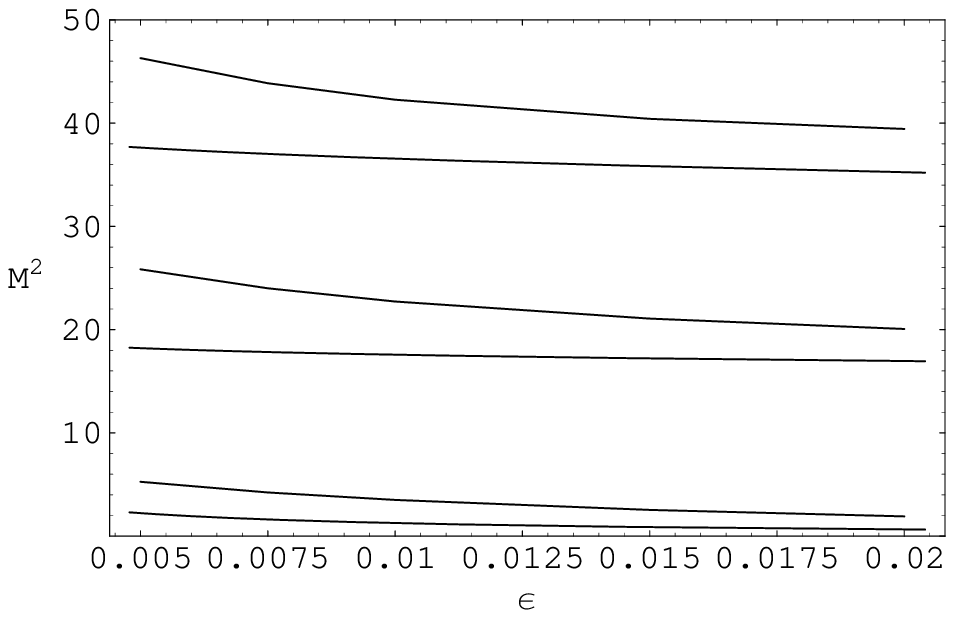}
%\centerline{\epsfxsize=2.9in\epsfbox{kim_mephi_lep.ps}}
\caption{The toy model (left) and $SYM_2$ with adjoint
scalars (right).}
\end{center}
\label{fig3}
\end{figure}

In the summary, reliable extrapolation of LCD data
requires very large cutoffs, while the IE do not
 have this problem. On the other hand, LCD samples better
multiparton Fock states which is difficult to achieve with
the IE. In that sense the two methods are complementary and
only comparison of both can provide unbiased information about
many-parton phenomena in the continuum limit of quantum field theories
including QCD.
\section*{Acknowledgements} All results presented in
Sects. 2 -- 5
were obtained in collaboration with G. Veneziano. I would like to
thank him for the continuous and enlightening discussions. I also
thank the Organizers for their hospitality.

\end{document}